\documentclass[10pt,aps,pre,superscriptaddress,onecolumn]{article}
\usepackage[english]{babel}
\usepackage{siunitx}  
\usepackage{enumitem}
\usepackage{graphicx, color}
\usepackage{authblk}

\usepackage{lineno}

\usepackage{hyperref}
\hypersetup{
 colorlinks=true,
 linkcolor=blue,
 filecolor=magenta, 
 urlcolor=blue,
 citecolor = blue,
 pdftitle={Overleaf Example},
 pdfpagemode=FullScreen,
 }
\usepackage[backend=biber,sorting=none,style=nature,citestyle=nature, maxbibnames=999,isbn=false]{biblatex}
\newcommand{\I}{\mathrm{i}}

\newcommand*{\defeq}{\mathrel{\vcenter{\baselineskip0.5ex \lineskiplimit0pt
                     \hbox{\scriptsize.}\hbox{\scriptsize.}}} =}

\graphicspath{{Figures/}}

\title{Wideband dispersion-free THz waveguide platform}

\author{David Rohrbach}
\author{Bong Joo Kang}
\author{Elnaz Zyaee}
\author{Thomas Feurer}
\affil{Institute of Applied Physics, University of Bern, Sidlerstrasse 5, 3012 Bern, Switzerland}
\date{}                     
\begin{document}
\maketitle

\begin{abstract}
We present an integrated THz spectroscopy and sensing platform featuring low loss, vacuum-like dispersion, and strong field confinement in the fundamental mode. Its performance was characterized experimentally for frequencies between 0.1~THz and 1.5~THz. While linear THz spectroscopy and sensing gain mostly from low loss and an extended interaction length, nonlinear THz spectroscopy would also profit from the field enhancement associated to strong mode confinement. Moreover, the vacuum-like dispersion allows for a reshaping-free propagation of broadband single- to few-cycle pulses in gas-phase samples or velocity matching between THz pump and visible to infrared probe pulses. Our platform is based on a metallic structure and falls in the category of double ridged waveguides. We characterize essential waveguide properties, for instance, propagation and bending losses, but also demonstrate junctions and interferometers, essentially because those elements are prerequisites for integrated THz waveform synthesis, and hence, for coherently controlled linear and nonlinear interactions.
\end{abstract}

A variety of THz applications, for instance, communication \cite{Piesiewicz2008, Federici2010, Kleine-Ostmann2011, Chen2019}, particle acceleration \cite{Fabianska2014, Nanni2015, Zhang2018, Othman:19, Snively2020, Rohrbach2021, Baek2021a}, or spectroscopy \cite{Fleischer2011, Fleischer2012, Hwang2015, Lu2016a} and sensing \cite{Ren2019, Akter2021} would benefit greatly from a versatile integrated platform. In the past, a number of different approaches have been proposed \cite{Mitrofanov2011, Atakaramians:13}. Among them different dielectric waveguide geometries showing low attenuation \cite{Nielsen2009, Lai:09, Bao2015a} and potentially strong mode confinement \cite{Nagel:06}. Interestingly, modern three-dimensional printing technologies allow for relatively simple fabrication of integrated dielectric THz waveguide structures \cite{Weidenbach2016_wg, Mavrona:21}. Their main drawback is the non-negligible mode dispersion which severely hampers broadband signal transmission inasmuch as group velocity dispersion leads to strong pulse broadening. Group velocity dispersion is less of an issue in metallic waveguides as the majority of the mode propagates in air. For instance, parallel metal plate waveguides show comparable little mode dispersion \cite{Mendis:01, Mendis2001}, however, the mode is confined only in one dimension, which limits the achievable field enhancement. Nevertheless, efficient free-space coupling \cite{Kim:10, Theuer2011, Shutler2012, Gerhard2012} and even fundamental building units of integrated circuits, such as T-junctions \cite{Pandey:10, Reichel2016}, were demonstrated. Similarly, metal wire based waveguides exhibit low attenuation and low dispersion \cite{Wang2004, Astley2010}, but suffer from high bending losses, modest mode confinement, and generally low free space coupling efficiency. Better performance is achieved for two parallel metal wires \cite{Mridha2014, Markov:14}, however, such structures are fragile and often the wires must be encapsulated with a dielectric material to mechanically stabilize the geometry, which in turn increases the mode dispersion and/or the attenuation. A very similar performance is found for slot-line waveguides \cite{Pahlevaninezhad:11}, which are especially well suited for very compact integrated circuits \cite{Sengupta2018a, Smith:21}. If no supporting substrate is present, the structure essentially becomes a metallic slit waveguide \cite{Wachter2007}, which is of special interest since it combines a low group velocity dispersion with a low attenuation and a strong mode confinement \cite{Smith2014, Peretti2018} across a wide range of frequencies. Slit waveguides consist of two parallel metal ridges separated by an air gap, with the ridge width and the air gap being of similar or smaller size than the free space wavelength. Such structures have been fabricated by milling slits in silicon wafers and by subsequent metal coating \cite{Wachter2007}, or by carefully aligning two thin metal foils \cite{Baek2021a}. 

Here, we demonstrate a compact and robust yet flexible platform essentially using a slit waveguide as the central motif. It features low propagation and bending losses and allows for active THz pulse shaping, and hence for fully integrated, coherently controlled spectroscopy. In order to guarantee efficient coupling of free space single-cycle THz pulses, we present a judiciously designed horn antenna.

\section{Waveguide design}

\begin{figure} [ht!]
\centering
\includegraphics[width=\columnwidth]{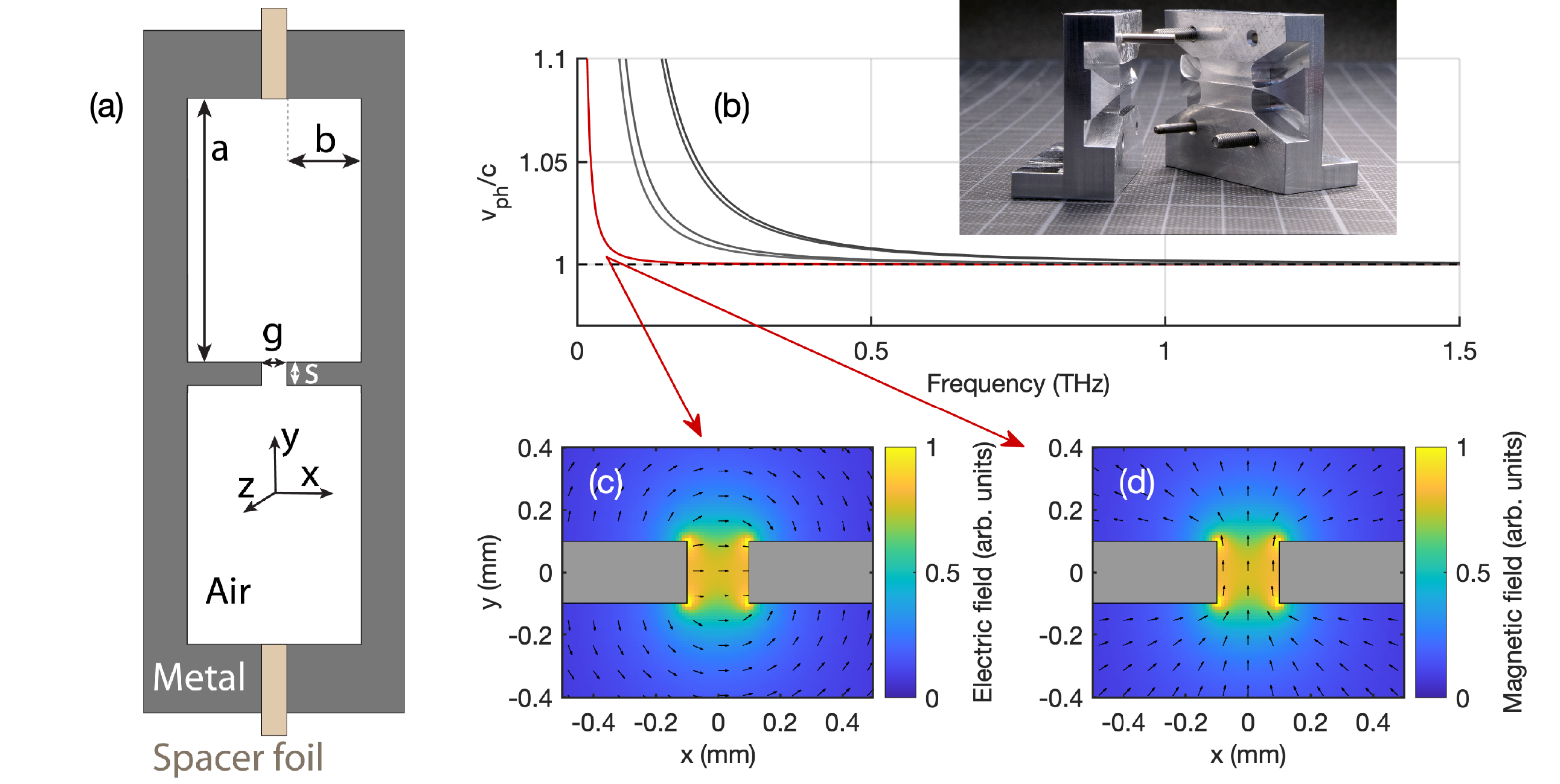}
\caption{\label{fig:ModeDispersion_Pattern} (a) Schematic cross-section of the double-ridged waveguide. (b) Waveguide dispersion of the lowest five Eigenmodes versus frequency. The inset shows a photograph of the manufactured waveguide structure with a 20~mm long straight channel including in- and out-coupling antennas. (c) Electric and (d) magnetic field distribution of the fundamental mode in the transverse $xy$-plane.}
\end{figure}

Figure~\ref{fig:ModeDispersion_Pattern}~(a) shows the cross-section of the proposed double-ridged waveguide. It consists of two mirror symmetric metal parts separated by a metallic spacer foil, which defines the gap between the two central ridges. In the limit of $a/s \gg 1$ and $b/g \gg 1$ the geometry resembles essentially that of a slit waveguide with the fundamental mode being confined to the gap volume. Our platform is not only robust mechanically, it also allows for integrating coupling elements, either to free space or to other waveguides, or for fabricating fully integrated functional elements, such as power splitters, filters, resonators, or interferometers. 

Here, we used numerical simulations to identify a geometry that features close to zero dispersion and low propagation and bending losses combined with strong mode confinement over the frequency range of interest of \SI{0.1}{THz} to \SI{1}{THz}, keeping in mind that the aspect ratio $b/s$ is limited by the manufacturing technique. For mode confinement, both $g$ and $s$ should in general be similar or smaller than the corresponding THz free space wavelength. The optimized parameters were found to $b = \SI{1}{\mm}$, $s = \SI{0.2}{\mm}$, and $a = \SI{5}{\mm}$. Note, that the gap size $g$ is given by the thickness of the spacer foils and was varied between $g = \SI{30}{\um}$ and $\SI{200}{\um}$. The inset in Fig.~\ref{fig:ModeDispersion_Pattern}(b) shows a photograph of the two mirror-symmetric parts for a \SI{20}{mm} long straight waveguide including the in- and out-coupling antennas at both ends. 

The simulated waveguide dispersion relation with the first five Eigenmodes is shown in Fig.~\ref{fig:ModeDispersion_Pattern}(b) for $g = \SI{200}{\um}$. The fundamental mode, which resembles a quasi-TEM mode with a cutoff frequency below \SI{10}{GHz}, exhibits a phase velocity (red curve) coinciding with the speed of light in vacuum (dashed horizontal line) down to about \SI{0.1}{THz}. The normalized transverse electric and magnetic field distribution of the fundamental mode are displayed in Figure~\ref{fig:ModeDispersion_Pattern}(c) and (d), with the color-coded amplitude and the arrows indicating direction of the respective field vectors. Note that the spatial mode profile is approximately frequency independent. While the electric field in the gap area is predominantly polarized along the $x$-direction, the magnetic field is mostly parallel to the $y$-direction.

\section{Fundamental mode characterization}

The experimental setup to characterize different waveguides is similar to the one reported in reference \cite{Gerhard2012}, with details given in the supplemental document. Briefly, single-cycle THz pulses are generated and detected by photo-conductive antennas, and the source is imaged to the detector via two pairs of aspherical lenses with focal lengths of \SI{100}{mm} and \SI{50}{mm}. The waveguides are positioned between the two central lenses and are equipped with input and output couplers for efficient coupling. Inspired by the cylindrical tapered couplers used in combination with parallel plate waveguides \cite{Gerhard2012}, we designed similar structures with a cylindrical tapering with a radius of $\SI{13}{\mm}$ in both transverse coordinates and a total length of \SI{10}{mm}. Similar geometries were successfully used for local field enhancement \cite{Zhan2011, Iwaszczuk2012}, but to the best of our knowledge never in combination with a slot waveguide structure. We simulated coupling efficiencies up to 80\% for the frequency range between \SI{0.35}{THz} to \SI{1}{THz}, which is in reasonable agreement with the experiments (see supplemental document for details). While coupling for lower frequencies is limited by reflection losses at the waveguide entrance, coupling for higher frequencies tends to be less efficient due to the smaller free space mode size.

\begin{figure} [ht!]
\centering
\includegraphics[width=1\columnwidth]{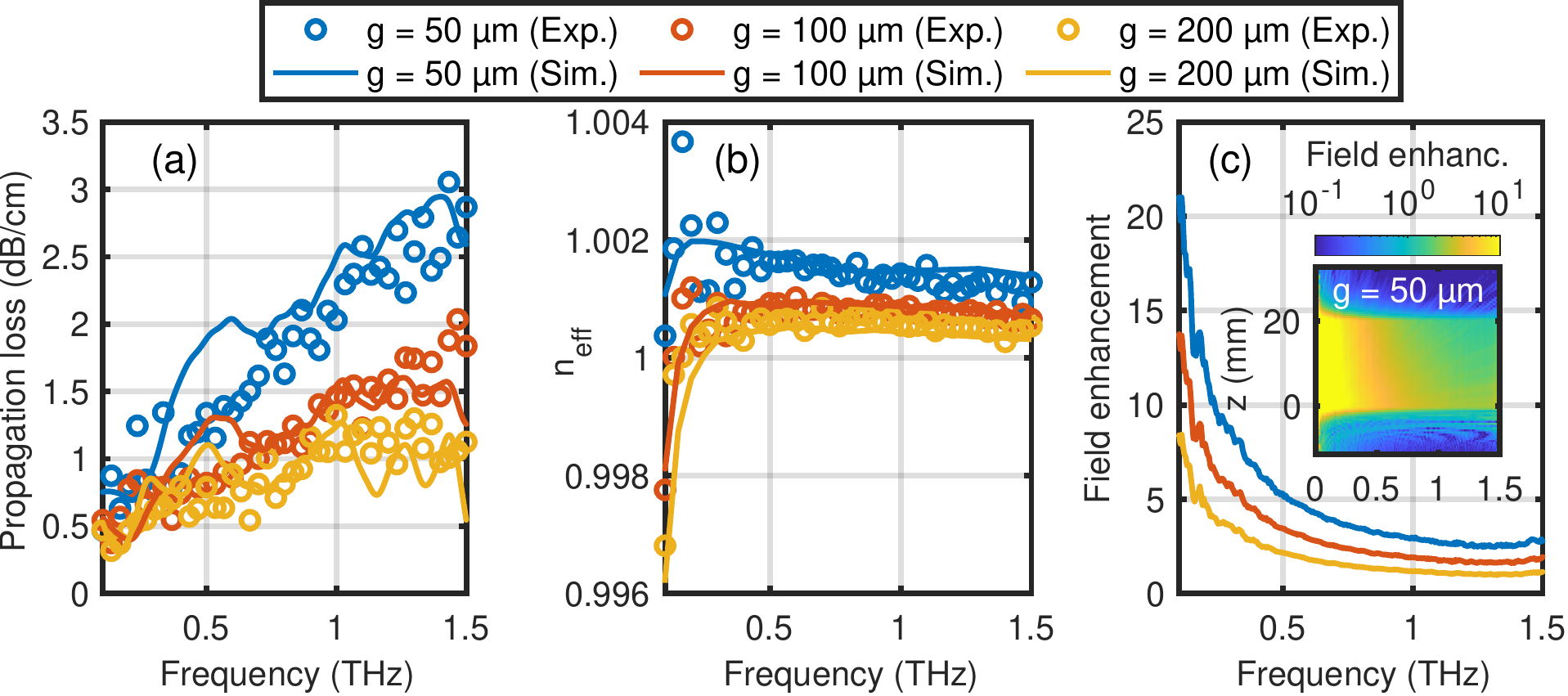}
\caption{\label{fig:StraightWG_losses} (a) Propagation loss, (b) effective refractive index, and (c) field enhancement at $z = \SI{0}{\mm}$ as a function of frequency for three different gap sizes of $\SI{50}{\um}$ (blue), $\SI{100}{\um}$ (red), and $\SI{200}{\um}$ (yellow). Open circles corresponds to experimental data and solid curves to simulations. The inset in (c) shows the simulated color-coded field enhancement as function of frequency and position along the middle of a 20~mm long waveguide with a gap of \SI{50}{\um}.}
\end{figure}

Figure~\ref{fig:StraightWG_losses} shows propagation loss, effective refractive index, as well as field enhancement of the fundamental mode as function of frequency for three different gap sizes of $\SI{50}{\um}$ (blue), $\SI{100}{\um}$ (red), and $\SI{200}{\um}$ (yellow). Generally, we find excellent agreement between measured (open circles) and simulated results (solid curves). The propagation loss increases with frequency, is larger for smaller gap sizes, and is governed mostly by Ohmic losses in the metal. With values on the order of a few \SI{}{dB \per \cm}, the propagation power loss is comparable to other THz waveguides, for instance, microstrip lines. Simulations suggest that lower attenuation can be achieved for metals with higher conductivity and should ideally vanish for superconducting materials, but in practice will be limited by scattering at waveguide imperfections. Figure~\ref{fig:StraightWG_losses}(b) shows that the effective index of refraction deviates at most by about $10^{-3}$ from the vacuum value down to \SI{0.1}{THz}. The deviation increases as the gap becomes smaller, and a similar behavior is found for surface plasmon polaritons on metal wires \cite{Wang2006} or in parallel plate waveguides \cite{Takeshima2013}. The variations of $n_{eff}$ between \SI{0.2}{THz} to \SI{1.5}{THz} are less than 0.2\%. This exceptional low waveguide dispersion is of utmost relevance for THz-pump optical-probe spectroscopy of gas-phase samples ensuring not only a reshaping-free propagation of single- or few-cycle THz pulses but also a velocity matching between an optical to infrared probe and the THz pulse. The rapid decrease for frequencies below \SI{0.2}{THz} is related to the mode dispersion as shown in Fig.~\ref{fig:ModeDispersion_Pattern}(c) and could be shifted towards lower frequencies by increasing the distance $a$ between the gap and the sidewalls. 

Since the transverse mode profile of the fundamental is smaller than the diffraction limited free space focus for a THz pulse with the same spectral content, an appreciable electric and magnetic field enhancement results. Here, we define the field enhancement factor shown in Fig.~\ref{fig:StraightWG_losses}(c) in frequency domain as the ratio of the peak electric field amplitude in the waveguide over the peak electric field amplitude at a free space focus using a f\# = 1 focusing element. Note that the field enhancement takes into account the coupling efficiency on the input side. The field enhancement can only be inferred from simulations, since we have no means to measure the electric field directly inside the gap. Similar enhancement factors are found for the magnetic field. As for most slit-based structures, lower frequencies show a higher field enhancement. This is because they have a larger mode profile when focused in free space, resulting in a larger increase in electromagnetic energy density when the mode is squeezed in the quasi-TEM mode of the waveguide. The inset in Fig.~\ref{fig:StraightWG_losses}(c) shows the color-coded field enhancement as a function of frequency and along the propagation axis. Note that especially the low frequency components are already enhanced toward the end of the input coupler ($z < 0$). The field enhancement decreases along the propagation axis in essence due to propagation losses. At 0.5~THz we find a peak field enhancement of 5 and an effective field enhancement of 4.4 when averaged over a \SI{20}{mm} long waveguide for $g = \SI{50}{\um}$. Hence, a lithium niobate THz source sufficiently strong to produce an electric field strength of \SI[per-mode=symbol]{400}{\kV \per \cm} when focused in free space, can be enhanced to almost \SI[per-mode=symbol]{2}{\mega \V \per \cm} along the entire waveguide. Similarly, the corresponding magnetic field would increase from \SI{130}{\milli\tesla} to \SI{590}{\milli\tesla}.

\section{Integrated circuits}

\begin{figure} [ht!]
\centering
\includegraphics[width=1\columnwidth]{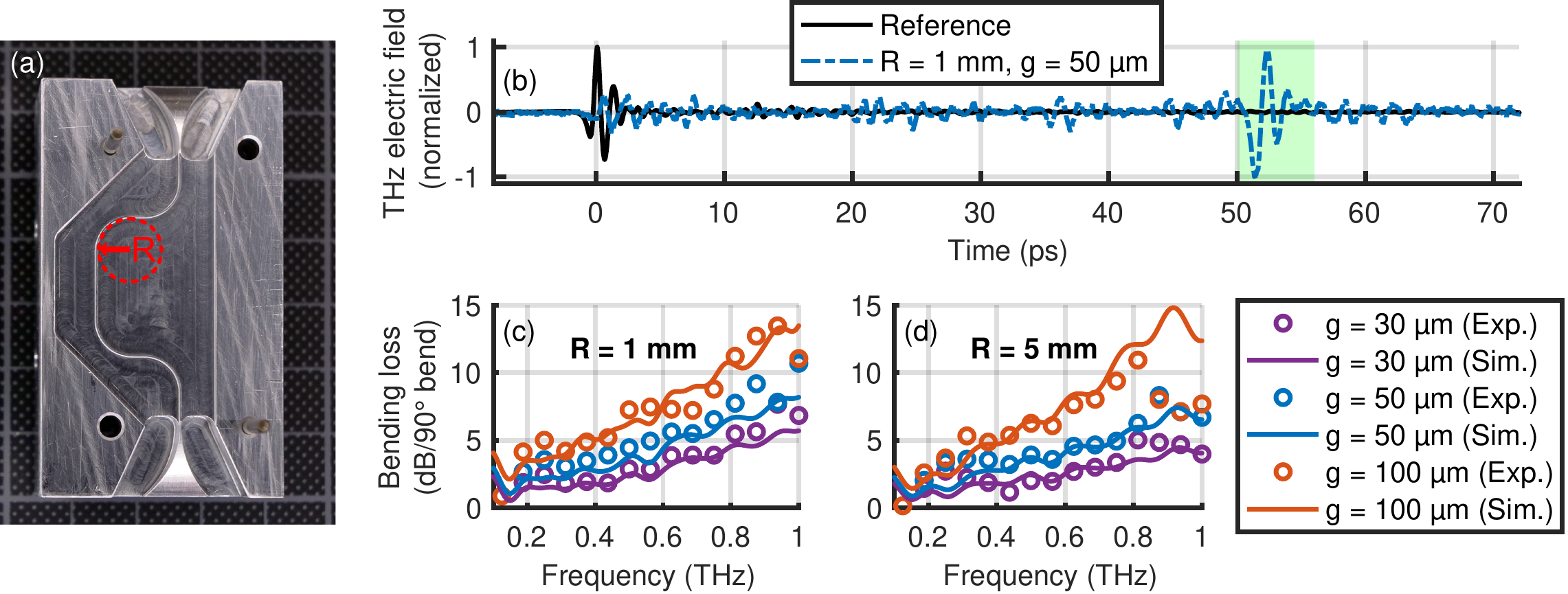}
\caption{\label{fig:BendingLosses} (a) Photograph of the bending structure with four 90$^\circ$ bends with an inner radius $R = \SI{5}{\mm}$. (b) Transmitted signal trough a structure with $R = \SI{1}{\mm}$ and $g = \SI{50}{\um}$ together with the normalized air-reference signal. (c) and (d) Extracted losses per 90$^\circ$ bend for different gap sizes and bending radii. Circles corresponds to experimental results and solid curves to simulations.}
\end{figure}

The strong confinement of the fundamental THz mode is especially beneficial for waveguide bends of small radii, on the order of millimeters, leading to low bending loss. Hence, our platform is suitable for developing highly sophisticated THz integrated circuits over a small footprint, for instance, power splitters or micro-ring based filters. In order to characterize the bending loss we use structures as the one shown in Fig.~\ref{fig:BendingLosses}(a) consisting of four 90$^\circ$ bends with radii of $R = \SI{1}{\mm}$ and $R = \SI{5}{\mm}$. Figure~\ref{fig:BendingLosses}(b) shows the measured transmitted electric field (blue dashed curve) as a function of time for $R = \SI{1}{\mm}$ and $g = \SI{50}{\um}$ together with the normalized reference signal (black solid curve). The green box indicates the time window in which we expect the signal guided around the four bends to appear. Compared to a straight waveguide the four bends increase the total waveguide length corresponding to a 51~ps time delay. The smaller signal contributions before and after the main guided signal are either due to scattering or leaking out of the waveguide. Figure~\ref{fig:BendingLosses}(c) and (d) show the losses per 90$^\circ$ bend as a function of frequency, for different gap sizes and bending radii. Generally, we find excellent agreement between experimental (open circles) and simulated results (solid curves). There is essentially no difference in performance when decreasing the radius from \SI{5}{\mm} to \SI{1}{\mm}. Even though a bending radius of $R=\SI{1}{\mm}$ is smaller than the free-space wavelength for frequencies below \SI{0.3}{THz}, we observe efficient guiding. Higher frequencies generally suffer from higher bending losses due to leakage. For gaps larger than \SI{100}{\um} no signal guiding was observed, but for smaller gaps guiding improves irrespective of frequency. Hence, there is a trade-off between bending loss and propagation loss, which can be optimized by adjusting the gap size.


\begin{figure} [ht!]
\centering
\includegraphics[width=1\columnwidth]{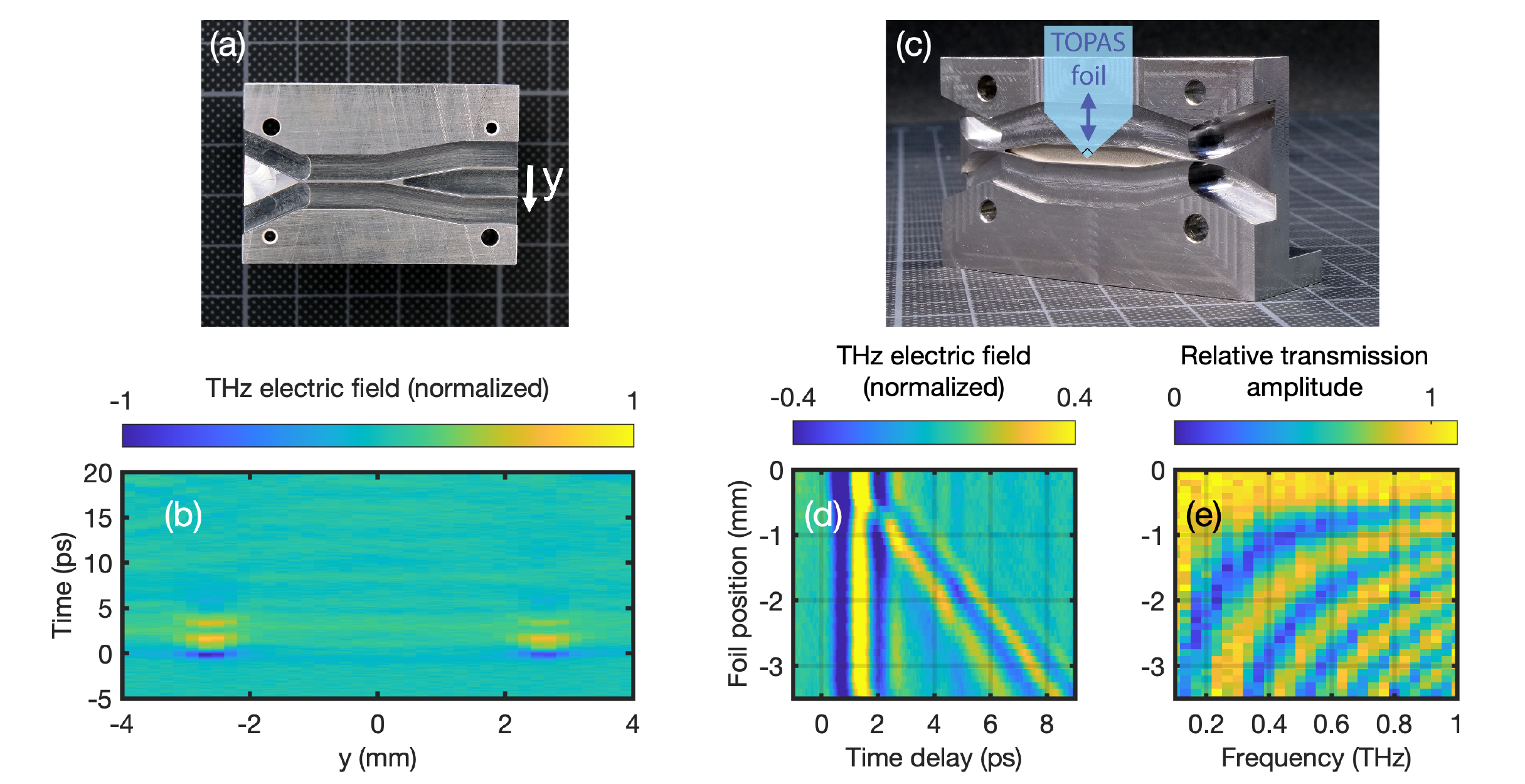}
\caption{\label{fig:Splitter_and_Retero} (a) Photograph of the Y-junction with in-coupler. (b) Measured electric field as a function of $y$-position and time. (c) Photograph of Mach-Zehnder interferometer based on two identical Y-junctions with in- and out-coupler. Position and geometry of the TOPAS foil are indicated by the blue shape. (d) THz electric field at the output of the interferometer for different foil position and (e) corresponding spectra.}
\end{figure}

THz waveform synthesizers for coherent control of linear and nonlinear interactions with matter are often realized via spectral phase and/or amplitude shaping. Integrated versions of such devices require as fundamental building blocks power splitters and there are a number of different options ranging from a standard Y-junction to a $1 \times 2$ rib-directional coupler, a parabolic-shaped structures, or a $1 \times 2$ arrow-2D directional coupler to name but a few \cite{Llobera2002_Thesis}. Here, we fabricated a Y-junction and based on that a Mach-Zehnder interferometer as a simple but fundamental waveform synthesizer. A photograph of the Y-junction is shown in Fig.~\ref{fig:Splitter_and_Retero}(a). The branching angle is kept as low as $20^\circ$ to minimize losses and the distance between the two parallel output channels is \SI{5.2}{\mm}. Figure~\ref{fig:Splitter_and_Retero}(b) shows the color-coded THz electric field as a function of time and $y$-position measured at the edge of the Y-junction structure using a near-field detection unit. Indeed, we find that the THz pulse is split in two replica with similar amplitude and time dependence. Small deviations from the designed symmetric splitting ratio of 50:50 are probably due to minor defects at the channel ends leading to different detection efficiencies. Note that the splitting ratio can be tuned by adjusting the branching angles of the two channels.


A photograph of the THz interferometer structure is shown in Fig.~\ref{fig:Splitter_and_Retero}(c). Basically, it consist of two mirror symmetric Y-junctions with two parallel waveguide sections in between. The separation of the two parallel waveguide sections is \SI{3}{\mm}. In order to delay the replica in the upper interferometer arm with respect to the other, we mount a 90$^\circ$-tip made of a \SI{80}{\um} thick dielectric foil on a linear stage moving the foil in $y$-direction in and out of the \SI{100}{\um} gap. The further the foil is moved in the gap, the larger the delay. The selected foil material, the cyclic olefin copolymer TOPAS, ensures low THz absorption \cite{Castro-Camus2020}. Figure~\ref{fig:Splitter_and_Retero}(d) shows the color-coded THz electric field after the interferometer as a function of time and foil position. The foil was moved in steps of \SI{100}{\um} for a total distance of \SI{3.5}{\mm}. At position \SI{0}{\mm} the foil is ineffective and the two replica superimpose without any delay between them. For positions smaller than \SI{-0.5}{\mm} one replica is delayed with respect to the other and we find a maximum delay of about \SI{6}{ps} at \SI{-3.5}{\mm}. Note that the delayed signal is slightly less intense due to Fresnel reflections and absorption. In addition, we observe an increasing dispersion as we move the TOPAS foil further in and simulations suggest this effect to be due to the air-gap between the TOPAS foil and the metal structure \cite{Mendis:06}. Figure~\ref{fig:Splitter_and_Retero}(d) shows the normalized color-coded THz spectra versus frequency and foil position and we observe the typical interference pattern for two time delayed THz pulses with the spectral fringes becoming more dense as the time delay increases. Well-controlled time delays between two or three phase-coherent THz pulses are essential for 2D-THz spectroscopy. Here for instance, the interferometer output can be coupled to a waveguide containing the sample and the nonlinear signal exiting the waveguide is sent to an electro-optic crystal for detection. For a three THz pulse interaction, the interferometer can be easily modified to produce an adjustable three pulse sequence.

\section{Conclusion}
We have demonstrated a versatile integrated THz platform based on waveguides featuring low bending and propagation losses of a few dB/cm, vacuum-like dispersion, and field enhancement factors of up to 5. Above 0.1~THz its performance is essentially frequency independent such that single- or few-cycle THz pulses, like those produced by optical rectification, propagate along the waveguide without reshaping. The low losses combined with the long interaction length and the field enhancement boost any nonlinear THz light matter interaction by orders of magnitudes. Moreover, the close to one effective refractive index of the fundamental mode allows for velocity matching between a THz pump and a visible to infrared probe pulse and hence for THz-pump optical-probe spectroscopy over the entire waveguide length. In addition, we have demonstrated a power-splitter and based on that an interferometer for THz waveform synthesis. Hence, this platform can be used as a fully integrated spectroscopy system for coherently controlled linear or nonlinear spectroscopy.

\section{Methods}

\subsection{Waveguide manufacturing}
All waveguide structures except for the THz interferometer were fabricated from aluminium using standard CNC-machining and subsequent lapping. For the Y-splitter, the minimum inner radius at the splitting region was set to 0.6~mm. To further improve the performance we use electrical discharge machining for the THz interferometer where the inner curvature radius reaches 0.2~mm. For this manufacturing technique we used \textit{steel 1.4034} instead of aluminium. 

\subsection{Evaluation}
\label{ssec:evaluation}
We consider the transmitted THz field through straight waveguides with different lengths $L$ of $\SI{20}{\mm}$, $\SI{40}{\mm}$ and $\SI{80}{\mm}$. Similar as in \cite{Mendis:01} we assume single-mode propagation such that the detected output electric field $E_{out}$ and the reference electric field $E_{ref}$ (i.e., without the waveguide structure) is given by 
\begin{equation}
\label{eq:modepropagation}
    E_{out}(\omega) = E_{ref}(\omega) \; T(\omega) \; C(\omega)^2 \exp\left(-\I [k_z -k_0] L\right) \exp\left(-\alpha L\right),
\end{equation}
where $\omega/(2 \pi)$ is the frequency, $T(\omega)$ is the total transmission coefficient taking into account reflections, $C(\omega)$ is the amplitude coefficient which is assumed to be the same at the entrance and the exit, $\alpha$ is the amplitude attenuation coefficient, $k_z$ is the $z$-component of the wavevector and $k_0 \defeq \omega/c$. By measuring the transmission through two waveguides of different lengths we can extract the propagation coefficient $\alpha$ and the wavevector $k_z$.

\section{Funding}
This project was funded by the Swiss National Science Foundation (SNSF) under grant no. 200020-178812.

\section{Acknowledgments}
The authors would like to cordially thank Adrian Jenk for manufacturing the waveguide parts and Rhoda Berger for taking the photographs. 

\section{Disclosures}
The authors declare no conflicts of interest.


\section{Supplementary Information}

\subsection{Experimental setup}
\begin{figure} [ht!]
\centering
\includegraphics[width=0.8\columnwidth]{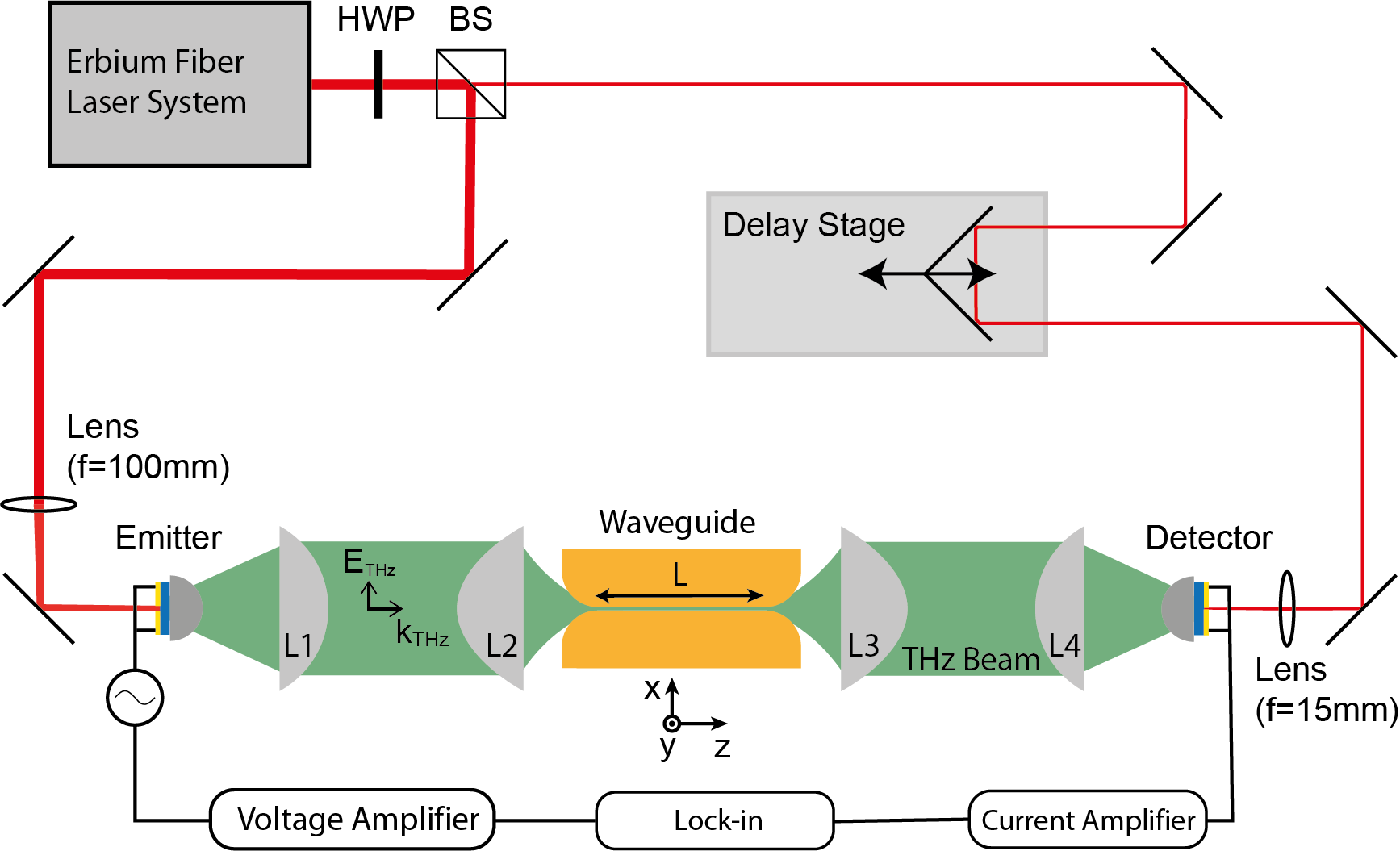}
\caption{\label{fig:SchematicTDSSetup} Schematic of the experimental setup: THz pulses are emitted and detected by two PCAs optically gated by a femtosecond laser pulse. The path length of the detector pulse can be adjusted by a delay stage.}
\end{figure}

A schematic of the setup is shown in Fig.~\ref{fig:SchematicTDSSetup}. It is based on a femtosecond laser \textit{Femto Fiber Pro} by \textit{Toptica}, with a pulse duration of less than 100~fs, a repetition rate of 80~MHz and an average output power of 130~mW at 780~nm. At the polarizing beam splitter (BS) the beam is split in a pulse for the emitter and the detector. Approximately 10~mW are transmitted through the BS and are used for the detection. The remaining part is used for THz generation at a photoconductive antenna (PCA). It is a so-called broad area interdigital PCA \textit{iPCA-21-05-1000-800-h} from \textit{Batop}.
As a bias voltage of the emitter PCA the 300~mV output voltage from the lock-in amplifier \textit{HF2LI} from \textit{Zurich Instruments} is increased to 15~V by a \textit{Falcon Systems} amplifier. The emitter PCA is orientated such that the THz pulse is horizontally polarized. The PCA is glued on a hyperhemispherical silicon lens, which increases the escape cone of the THz pulses. 
The path length of the detector pulse can be adjusted by a home-built delay line which consists of a hollow roof prism mirror from \textit{Thorlabs} mounted on a linear translation stage \textit{Minislide MSQSD7} from \textit{Schneeberger} driven by a linear motor \textit{LM1483} from \textit{Faulhaber}. 
The detector PCA is a bow-tie antenna \textit{PCA-100-05-10-800} from \textit{Batop} including a hyperhemispherical silicon lens on top of the substrate. We detect the THz-induced current with a \textit{HF2TA Current Amplifier} from \textit{Zurich Instruments}. For a lock-in detection scheme, the bias voltage of the emitter PCA is modulated at 2~kHz.
The THz pulse is transported to the detector PCA by four aspheric THz lenses. The focal length of $L1$ and $L4$ is 100~mm while the focal length of $L2$ and $L3$ is 50~mm. To adjust for the different waveguide lengths, the third THz lens $L3$ was moved in $\pm z$-direction.

\subsection{Technical drawing}
\begin{figure} [ht!]
\centering
\includegraphics[width=0.8\columnwidth]{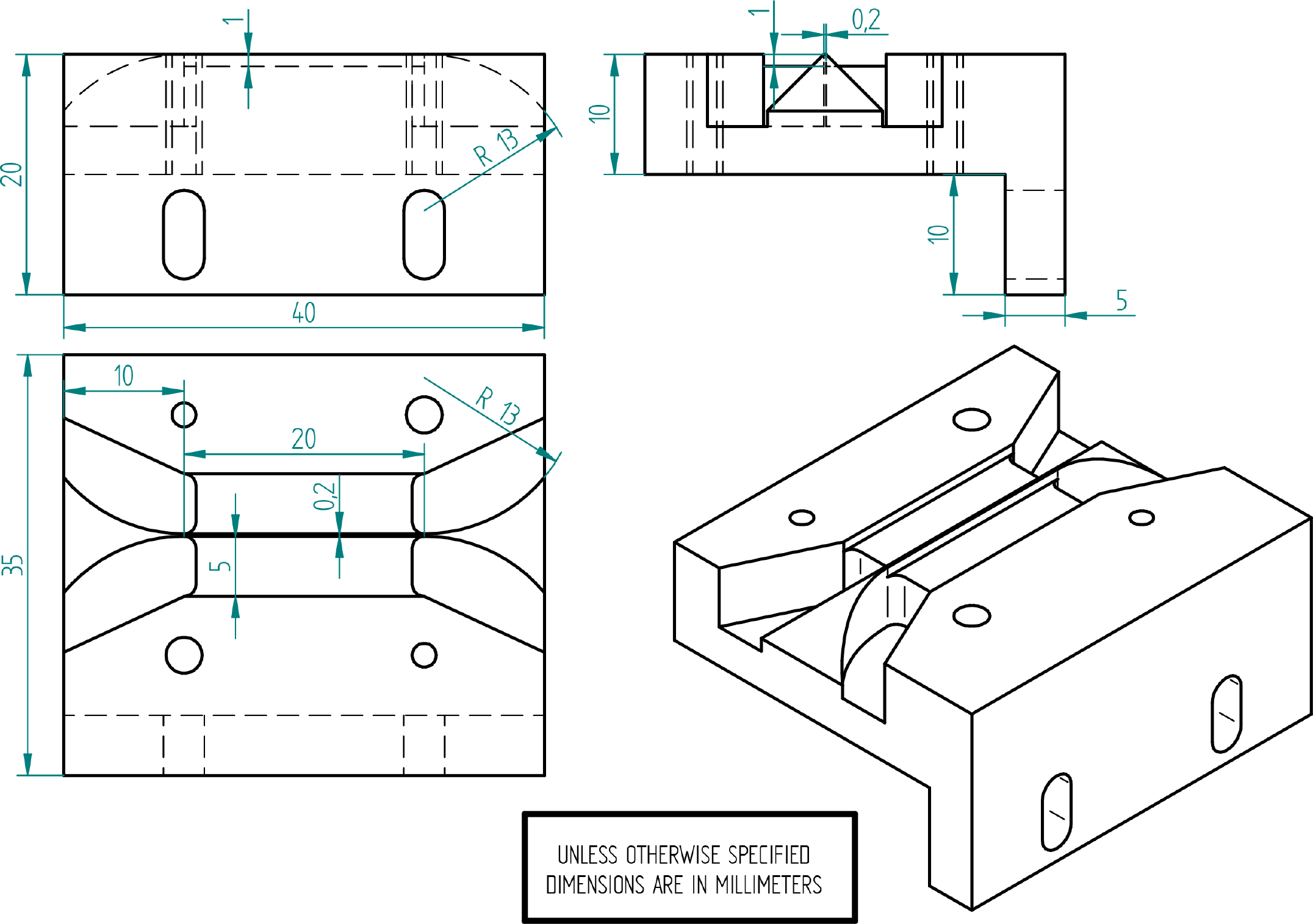}
\caption{\label{fig:Channel20mm_CAD} Technical drawing of one half of the 20~mm long waveguide structure with the in- and out-coupling structures. The most important dimensions are indicated in teal with units in millimeters.}
\end{figure}

Figure~\ref{fig:Channel20mm_CAD} shows the technical drawing of one half of the 20~mm long waveguide structures. The slit waveguide is built by assembling two mirror symmetric structures with a given gap size. Alignment holes are used for the accurate assembling of the two parts. 

\subsection{Coupling coefficient and misalignment effects}

\begin{figure} [ht!]
\centering
\includegraphics[width=0.8\columnwidth]{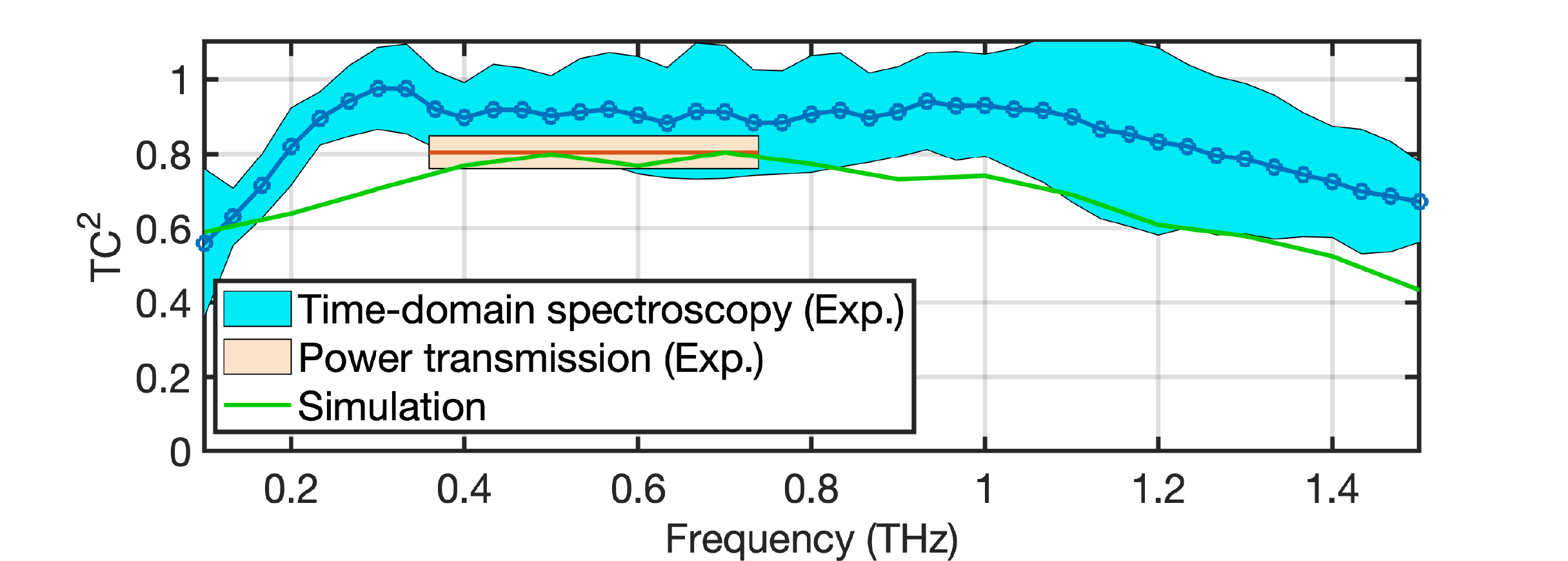}
\caption{\label{fig:StraightWG_coupling} Total coupling coefficient $T C^2$ for the optimal alignment based on time-domain spectroscopy measurements, direct transmission power measurements and numerical simulations.}
\end{figure}

Figure~\ref{fig:StraightWG_coupling} shows the extracted overall coupling coefficient $T C^2$ for a gap size of \SI{200}{\um}. Here, we compare the experimental data from the time-domain spectroscopy (TDS), direct power measurements in a high power THz setup (with center frequency at 0.5~THz and a power FWHM of 0.4~THz) and numerical simulations. The experimental data from the TDS and the simulations show that the optimal coupling is obtained for frequencies between 0.35~THz to 1~THz. The coupling of lower frequencies is limited by reflection losses at the waveguide entrance due to the larger modal size, while at higher frequencies the coupling to the TEM mode of the waveguide seems to be less efficient due to the reduced mode confinement. The best coupling efficiency based on the simulation is 80~\% and we even find somewhat larger values for the TDS experimental data, which we attribute to imperfect imaging of the reference signal \cite{Gerhard2012}. This is confirmed by the direct THz power transmission measurements in a high power THz setup where we find an averaged coupling coefficient around 80~\% for the considered THz frequency spectrum. The shaded area indicates the estimated uncertainty based on measurement of three different waveguide lengths. In order to reduce noise in the TDS-data, we apply a moving average along 0.2~THz. 

\begin{figure} [ht!]
\centering
\includegraphics[width=1\columnwidth]{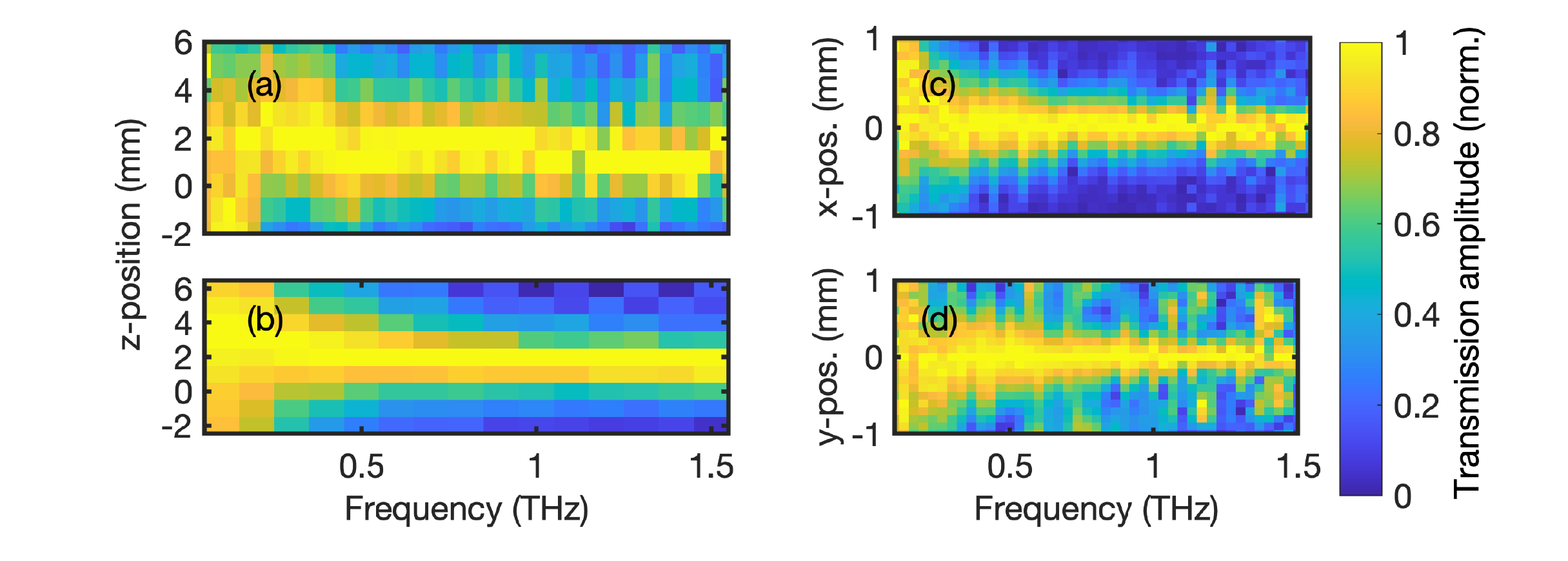}
\caption{\label{fig:StraightWG_alignment} Measured normalized transmitted spectra for a 80~mm long waveguide with a \SI{200}{\um} gap for different misalignment in $z$- (a), $x$- (c) and $y$-direction (d). In (b) the corresponding results from numerical simulations are shown for different $z$-alignments.}
\end{figure}

We measured the influence of slightly misalignment to the coupling efficiency using a 80~mm long waveguide with a fixed gap size of \SI{200}{\um}. Figure~\ref{fig:StraightWG_alignment}(a) shows the transmitted spectrum for different $z$-positions of the waveguide in steps of 1~mm. Note that here we adjusted also the position of the third THz lens $L3$ such that the distance from both lenses to the waveguide remained the same. The transmission spectrum is normalized to the peak transmission for a given frequency. The $z$-position at 0~mm corresponds to a waveguide alignment such that the THz focus plane is located at the transition between the coupling structure and the straight waveguide. Interestingly, the highest coupling coefficient is found for $z = \SI{2}{\mm}$, which corresponds to the focal plane located inside the coupler, in 2~mm distance to the waveguide beginning. In order to verify this finding, we performed numerical simulations and calculate the coupling coefficient based on the mode overlap integral of the electric fields at the waveguide output. The corresponding results are shown in Fig.~\ref{fig:StraightWG_alignment}(b) where we again find an optimal coupling for $z = \SI{2}{\mm}$.

Furthermore, we consider misalignment of the waveguide in $x$- and $y$-direction in steps of \SI{100}{\um} by a total of $\SI{\pm1}{\mm}$, as shown in Fig~\ref{fig:StraightWG_alignment}(c) and Fig~\ref{fig:StraightWG_alignment}(d) respectively. The coupling of lower frequencies seems to be less affected by misalignment than the higher frequency parts, where the waveguide must be aligned with \SI{100}{\um} precision in both lateral dimensions for optimal performance. Any asymmetry of the coupling efficiency with respect to $x = \SI{0}{\mm}$ or $y = \SI{0}{\mm}$ is attribution mainly to asymmetries in the THz setup. For the largest misalignment in $y$-direction, the THz pulse is also partially guided in the air region parallel to the waveguide channel, which explains the weaker decrease in the coupling efficiency for misalignment in $y$-direction compared to the $x$-direction.

\subsection{Details about numerical simulations}
Three-dimensional time-domain simulations of a focused THz pulse propagating through different waveguide geometries were performed in CST Microwave Studio. The simulation effort was reduced by applying symmetry conditions for the $xz$- and $yz$-plane. The boundary perpendicular to the $y$-direction was set as perfect electrical conductor, while in all other directions open boundary conditions were applied to eliminate artificial reflections. For the aluminium waveguide simulations we assumed a conductivity of $\SI{12}{\mega \siemens \per \m}$ based on fitting to the experimental results. The focused linearly polarized THz pulse was modelled with a transverse Maxwell-Gaussian beam profile \cite{Davis1979, Barton1989}. To account for the diffraction limited THz focusing, the beam waist was scaled inversely proportional with frequency and we set the beam waist to 0.5~mm at 0.6~THz.

\printbibliography


\end{document}